\documentclass[fleqn,usenatbib]{mnras}

\usepackage{newtxtext,newtxmath}
\usepackage[T1]{fontenc}
\usepackage{ae,aecompl}

\newcommand{\mr}{\mathrm}

\newcommand{\az}{$\langle z \rangle$}
\newcommand{\azm}{z_0}

\usepackage{graphicx}	% Including figure files
\usepackage{amsmath}	% Advanced maths commands
\usepackage{amssymb}	% Extra maths symbols

\usepackage{CJKutf8}

\title[FRB Redshift]{Statistical inference of the distance to ASKAP FRBs}

\author[D. Z. Li et al.]{
Dongzi Li\,(李冬子)$^{1,2,3}$\thanks{E-mail: dzli@cita.utoronto.ca},
Almog Yalinewich$^{1}$ and
Patrick C. Breysse$^{1}$
\\
$^{1}$Canadian Institute for Theoretical Astrophysics, University of Toronto, 60 St. George Street, Toronto, ON M5S 3H8, Canada\\
$^{2}$Department of Physics, University of Toronto, 60 St. George Street, Toronto, ON M5S 1A7, Canada\\
$^{3}$Dunlap Institute for Astronomy and Astrophysics, University of Toronto, 50 St. George Street, Toronto, ON M5S 3H4, Canada
}

\date{Accepted XXX. Received YYY; in original form ZZZ}

\pubyear{2019}

\begin{document}
\begin{CJK*}{UTF8}{gbsn}

\label{firstpage}
\pagerange{\pageref{firstpage}--\pageref{lastpage}}
\maketitle

\begin{abstract}
The distances to fast radio bursts (FRBs) are crucial for understanding their underlying engine, and for their use as cosmological probes.
In this paper, we provide three statistical estimates of the distance to ASKAP FRBs. 
First, we show that the number of events of similar luminosity in ASKAP does not scale as distance cubed, as one would expect, when directly using the observed dispersion measure (DM) to infer distance. 
Second, by comparing the average DMs of FRBs observed with different instruments, we estimated the average redshift of ASKAP FRBs to be $z\sim 0.01$ using CHIME and ASKAP,  
and $z\lesssim0.07$ using Parkes and ASKAP. 
Both values are much smaller than the upper limit $z\sim0.3$ estimated directly from the DM. 
Third, we cross-correlate the locations of the ASKAP FRBs with existing large-area redshift surveys,  
and see a 3\,$\sigma$ correlation with the 2MASS Redshift Survey and a 5\,$\sigma$ correlation with the HI Parkes All Sky Survey at $z\sim0.007$. 
This corresponds well with the redshift of the most likely host galaxy of ASKAP FRB 171020, which is at 
$z=0.00867$. 
These arguments combined suggest an extremely nearby origin of ASKAP FRBs and a local environment with accumulated electrons that contribute a DM of several hundred pc/cm$^3$, which should be accounted for in theoretical models.
\end{abstract}


%%%%%%%%%%%%%%%%%%%%%%%%%%%%%%%%%%%%%%%%%%%%%%%%%%

%%%%%%%%%%%%%%%%% BODY OF PAPER %%%%%%%%%%%%%%%%%%

\section{Introduction}
Fast radio bursts (FRBs) are short (few microseconds) radio signals occurring at seemingly random times and places in the sky. Their origin is unknown, and is currently an active field of research \citep[see][ and references therein for a review]{Katz2016FastAnswers, Ravi2017TheBursts, Katz2018FastBursts}. At this point, there are dozens of proposed FRB progenitor models \citep[as reviewed in][]{Platts2018ABursts}, motivating efforts to constrain the space of possible models.
 
One significant source of uncertainty regarding FRBs is their distance or redshift. Measurement of the dispersion measure (DM) places an upper bound on the redshift assuming all of the dispersion is due to the intergalactic medium (IGM). However, there is evidence to suggest that the actual redshift can be considerably lower than this bound. First, \citet{Luan2014PhysicalBurst} and \citet{Zhu2018TheMeasures} show that the scattering measure of observed FRBs cannot be explained by the IGM alone, which suggests that there is additional plasma along the line of sight which may contribute to the DM. Second, the host of the repeating FRB 121102 \citep{17Tendulkar} is found at a much lower distance than that inferred from the DM, as is the most likely host for FRB 171020 \citep{Mahony2018AFRB171020}. 
Currently, many analyses of the FRB population, luminosity function and detectability at high redshift explicitly or implicitly rely on the assumption that the DM is dominated by the IGM contribution \citep[e.g.][]{Shannon2018TheSurvey, Zhang2018FRBRedshifts, Luo18FRBluminosity}. 

In addition to settling this controversy, determining the distances of FRBs will allow us to calculate the luminosities and contribution of the host or circumburst environment to the DM, two important clues to the nature of the FRB engine. Moreover, the slope of the high-energy tail of the luminosity function will determine the detection rate of events emitted at high redshift. The abundant cosmological applications of FRBs all require a large sample of FRBs at a relatively high redshift (i.e.\,$z\gg 0.1$) \citep[e.g.][etc]{Zhou2014FastProbe, Munoz2018FindingMaps, Jaroszynski2018FastTests}. 

Due to their short duration, it is difficult to determine the detailed properties of FRBs. The DM and flux are measured for almost all FRBs, but other properties like the polarisation and rotation measure are only measured for a handful of events \citep{Petroff2014AFollow-up,  Masui2015DenseBurst, Keane2016TheBurst, Ravi2016TheBurst, Petroff2017ALatitude}. A rough sky position is measured for most FRB's, but the accuracy is not sufficient to pinpoint the emission to a particular galaxy. However, as we show in this paper, the localization may be good enough to associate FRBs with large-scale structure at different epochs. 

The paper is organized as follows: section\,\ref{sec:sta} provides two statistical inferences of the FRB distances from the DMs, based on the scaling of number counts with DM and comparison between different experiments; 
section\,\ref{sec:corr} correlates the ASKAP FRB positions with large scale structures in galaxy surveys; 
conclusion and discussions are in section\,\ref{sec:conclude}.
 
\section{Statistical Arguments for proximity}
\label{sec:sta}

In this section we present some some statistical arguments that support the idea that the  distance to FRBs is considerably lower than the upper limit from the DM. Note that when we refer to DM below we mean the DM$_\mr{excess}$, from which the Milky Way contribution has been subtracted. Also, the DM contribution from the host DM$_\mr{host}$ includes contributions from both the circumburst environment and the host galaxy interstellar medium. 

\subsection{Relation between Luminosity and Dispersion Measure}
% The notebook can be found here https://colab.research.google.com/drive/1xY_nClIy_Omu7KEEwIA92jg3LxeoWXWX

Let us assume for now that the contribution of the host to the DM is negligible. If this is the case, then the DM is solely due to the IGM, and therefore scales linearly with the distance to the host. If the distances are known, then one can readily calculate the bolometric luminosity from the observed flux. For a low enough redshift $z\ll1$, one can assume that space is Euclidean, and that star formation has not evolved considerably from local values. In this case the number of events of the same luminosity below a certain radius $r$ should increase as $r^3$. Since the DM is linear in the distance, then we also expect the number to increase with the DM cubed. 

\begin{figure}
  \includegraphics[width=\hsize]{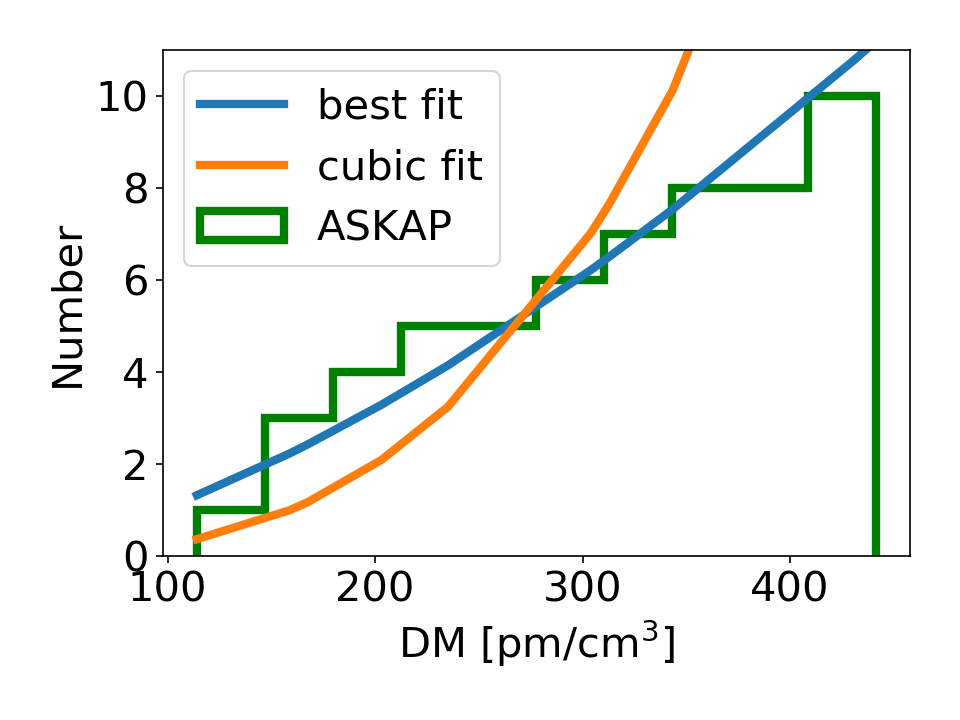}
  \caption{Cumulative number distribution of FRBs with similar hostless luminosities from ASKAP, as a function of DM (green curve). The hostless luminosities were calculated under the assumption that all of the dispersion is due to the intergalactic medium, so they are proportional to the product of the flux and DM squared. Best fit is obtained when the count scales as $DM^{1.5}$ (blue) rather than $DM^{3}$ (orange).
  }
  \label{fig:hostless}
  \end{figure}

To test this hypothesis, we require a population of FRBs with the same ``hostless" luminosity. We therefore choose ten FRBs in ASKAP, which, according to this conjecture, lie within a range of factor of two in luminosity, and plot the cumulative number count as a function of DM (corrected for the contribution of the milky way) in figure \ref{fig:hostless}. This figure shows that the number of FRBs scales more like $DM^{1.5}$ than $DM^3$ as would be predicted by the hostless model. We note that due to Poisson noise, this difference is not statistically significant. However, it is suggestive of an issue with the hostless, IGM-dominated model.

\subsection{Comparison between Different Instruments}
\begin{center}
\begin{table}
\begin{tabular}{|lllll|} 
\hline
 &N$_\mr{FRB}$  & S/S$_\mr{ASKAP}$  & $\overline{\mr{DM}}$ &$z_\mr{ASKAP}$ \\ \hline 
 ASKAP	&23	 &1  & $390\pm40$ & \\ %\hline
 CHIME	&13  & 50 & $470\pm 80$ &$0.01\pm0.01$ \\ %\hline
 Parkes		&22  & 50 & $840\pm 120$ & $\lesssim 0.07\pm0.02$ \\
 \hline
\end{tabular}
\caption{
%Statistics of FRBs detected by three majoy instruments.
N$_\mr{FRB}$: number of published FRBs; S/S$_\mr{ASKAP}$: sensitivity with respect to ASKAP; 
$\overline{\mr{DM}}$: average excess DM, the errors are the statistical fluctuations; 
$z_\mr{ASKAP}$: the inferred average redshift of ASKAP FRBs, the errors are propagated from the error of DM, which is a lower bound of the actual uncertainty.
}
\label{tab:dm}
\end{table}
\end{center}
In the previous section we argued for a low-redshift FRB model under the simplifying assumption that there is no host contribution to the DM.  We will now present another argument which takes the host dispersion into account.

%For most of the FRBs Detected today, the distance upper limit from DM suggest it come from redshift below 1. In the distance, the universe can be assumed to be Euclidean and the evolution effect can be small. 
Assume that the FRBs detected by different instruments are emitted from the same populations (and thus have similar statistical properties), and that the detection rate is flux rather than volume limited.  Under these assumptions, the average distance to the detected FRBs in Euclidean space will be proportional to the square root of the instrument sensitivity $\langle d \rangle \propto \sqrt{S}$. 
This relationship holds for any type of luminosity function (See Appendix \ref{app:lum} for a proof of this statement).
Therefore, 
\begin{equation}
\frac{ \langle \mr{DM}_\mr{IGM} (d_1) \rangle}{ \langle \mr{DM}_\mr{IGM} (d_2) \rangle}\approx\frac{\langle d_1 \rangle}{\langle d_2 \rangle}=\sqrt{\frac{\langle S_1\rangle}{\langle S_2 \rangle}},
\label{eq:IGM}
\end{equation}
where $\langle d_1\rangle$ and $\langle d_2\rangle$ are the average distances of the FRBs detected in two arbitrary instruments with sensitivities $S_1$ and $S_2$. For $z\lesssim 1$ the $\mr{DM}_\mr{IGM}$ is almost linearly proportional to the distance in the standard calculation \citep{04Inoue}.

The expected excess DM from the two experiments can be written as a sum of contributions from the host and the IGM,
\begin{align}
    \langle \mr{DM}(d_1) \rangle  = \langle \mr{DM}_\mr{host}(d_1) \rangle  + \langle \mr{DM}_\mr{IGM} (d_1) \rangle \\
        \langle \mr{DM}(d_2) \rangle  = \langle \mr{DM}_\mr{host}(d_2) \rangle  +\langle \mr{DM}_\mr{IGM} (d_2) \rangle
        %&\rightarrow{}\langle d^\prime\rangle-\langle d\rangle
\end{align}
For $z\ll1$, the host properties are expected to evolve only weakly, so 
\begin{equation}
	\langle \mr{DM}_\mr{host}(d_1) \rangle\approx\langle \mr{DM}_\mr{host}(d_2) \rangle.
	\label{eq:host}
\end{equation}
Eqs. \ref{eq:IGM}-\ref{eq:host} contain a total of four unknowns, so we can straightforwardly compute the average $\mr{DM}_\mr{IGM}$ for both telescopes, and therefore the mean redshifts.

We apply this calculation to FRBs detected by three instruments: Parkes, ASKAP and CHIME \citep{Amiri2018TheOverview}. 
The results are presented in Table.\,\ref{tab:dm}. We focus here as throughout this work on determining the redshifts of the ASKAP FRBs.  Comparing ASKAP and CHIME yields an average redshift for these objects of $z_\mr{ASKAP}=0.01\pm 0.01$, while the ASKAP/Parkes comparison gives $z_\mr{ASKAP}=0.07\pm0.02$.  The errors are propagated from the variance of the average DM, which is calculated by $\sigma_\mr{DM}/\sqrt{N-1}$, where $\sigma_\mr{DM}$ is the standard deviation of the observed DM. 

In addition to the statistical uncertainties quoted above, there may be systematic effects which affect this computation. 
%The minimum detectable peak flux density for FRBs is increasing faster for DM$<\sim600\,$\DMunit than greater than DM$>\sim1000$ \citep{18Patel}.
First, the instruments we are comparing observe in very different radio environments.
The radio frequency interference (RFI) in the vicinity of Parkes is much stronger than around ASKAP, 
reducing the chances of detecting low-DM FRBs (Figure 2 of \citealt{18Patel}). This RFI would systematically increase the value of $\langle z_\mr{ASKAP} \rangle$ when comparing Parkes to ASKAP. For this reason, we quote our Parkes/ASKAP estimate in Table \ref{tab:dm} as an upper limit.
CHIME also observes in a relatively high-RFI environment, but it also observes at much lower frequencies ($\sim$600 MHz for CHIME vs. $\sim$1.3 GHz for ASKAP), where the differential delay is 16 times larger given the same DM with ASKAP. So the CHIME sensitivity to low-DM FRBs is less affected.  This frequency difference, however, creates the potential for additional systematics, as the FRB luminosity may evolve with frequency. More generally, each experiment uses their own unique algorithms for searching for FRBs, which could bias the detection rates in different ways.  Therefore, the estimated $z_\mr{ASKAP}$ can have much larger uncertainties than the provided error bars. 
Nevertheless, the $z_\mr{ASKAP}$ estimated from both sets are an order of magnitude smaller than $z_\mr{ASKAP}\sim0.3$ obtained for IGM-dominated DMs.  This is consistent with the argument from the previous section.

\begin{figure}
  \includegraphics[width=\hsize]{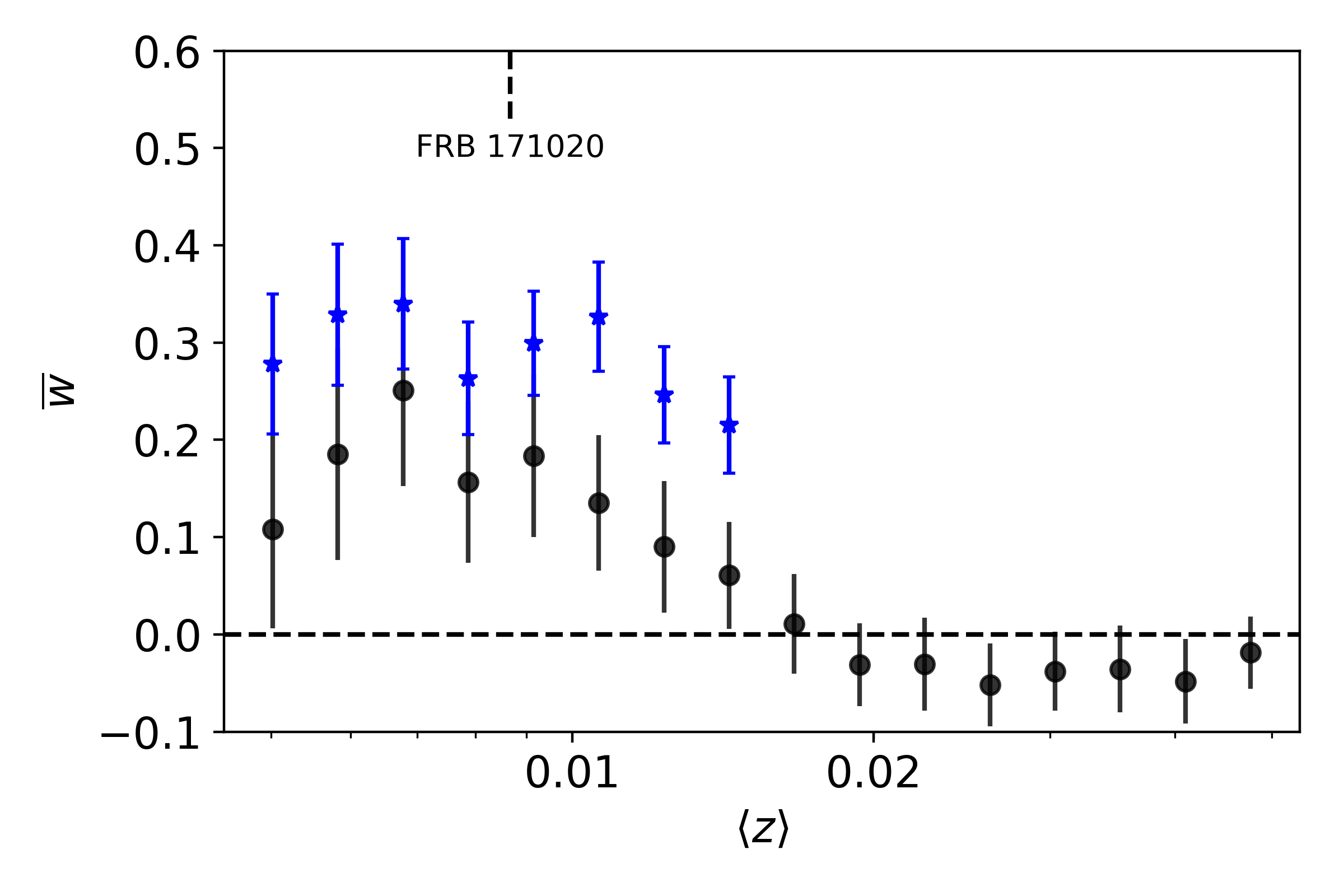}
  \caption{The correlation amplitude $\overline{w}(\azm)$ between ASKAP FRBs and galaxies detected in 2MRS (dark circles) and HIPASS (blue stars) as a function of redshift. Note that the redshift weightings of nearby points significantly overlap, so the error bars are highly correlated.
%The selected redshift bins are large, so there are significant overlap in redshift space between nearby points. 
      %The profile against redshift is largely related to the parameterized function we select for slicing, because the thickness of the slice is much larger than the difference of $\langle z\rangle$ for nearby points. 
 %The green and yellow bar indicate the ASKAP FRBs' average redshift estimated from their mean DM difference with the CHIME FRBs and Parkes FRBs respectively (the later is biased higher); 
  The dashed vertical line shows the redshift of the most likely host galaxy of the ASKAP FRB 171020\,\citep{Mahony2018AFRB171020}.  
    }
  \label{fig:totcorr}
  \end{figure}
  
 \begin{figure}
     \begin{minipage}{\linewidth}
        \center
  \includegraphics[width=\hsize]{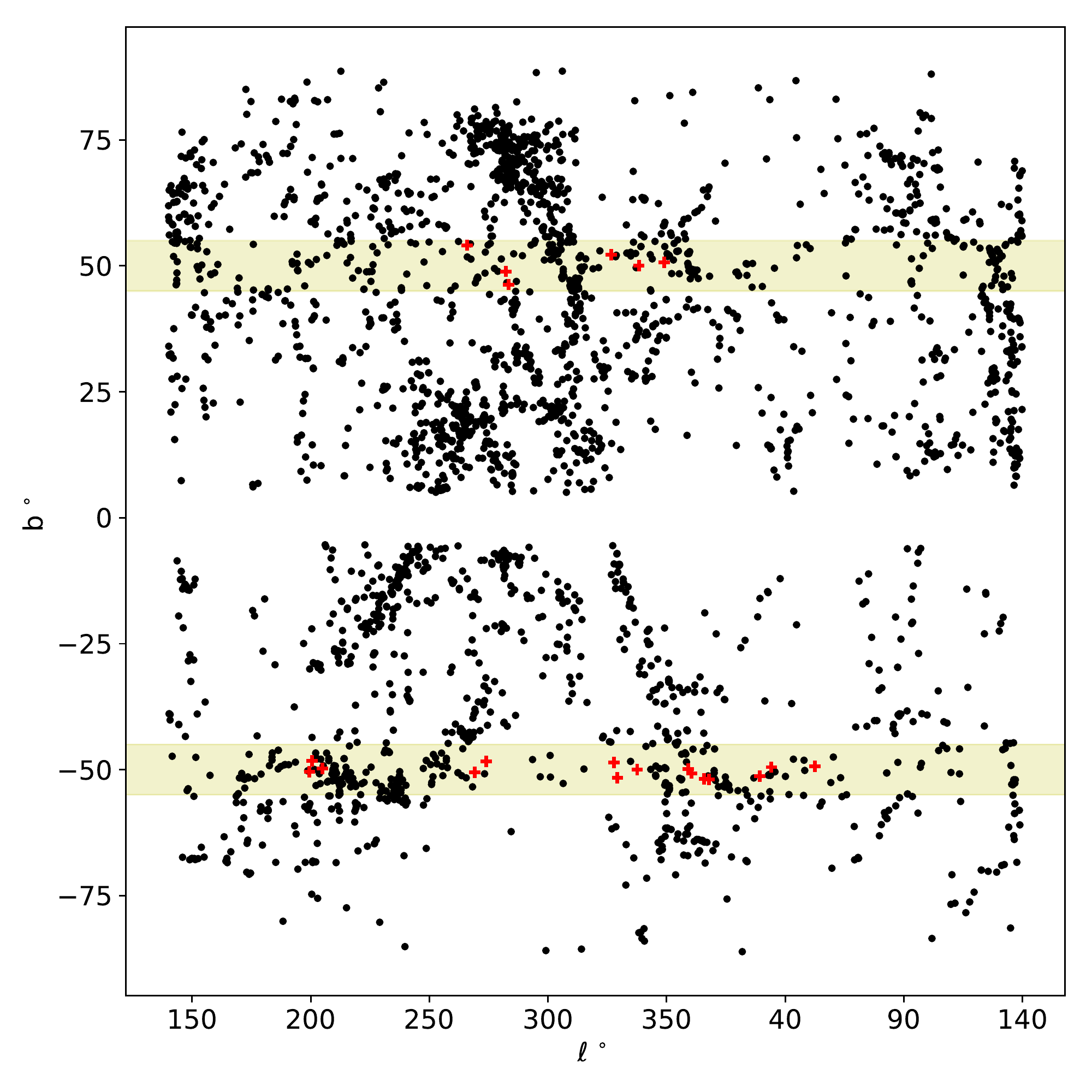}
  \vfill
    \includegraphics[width=\hsize]{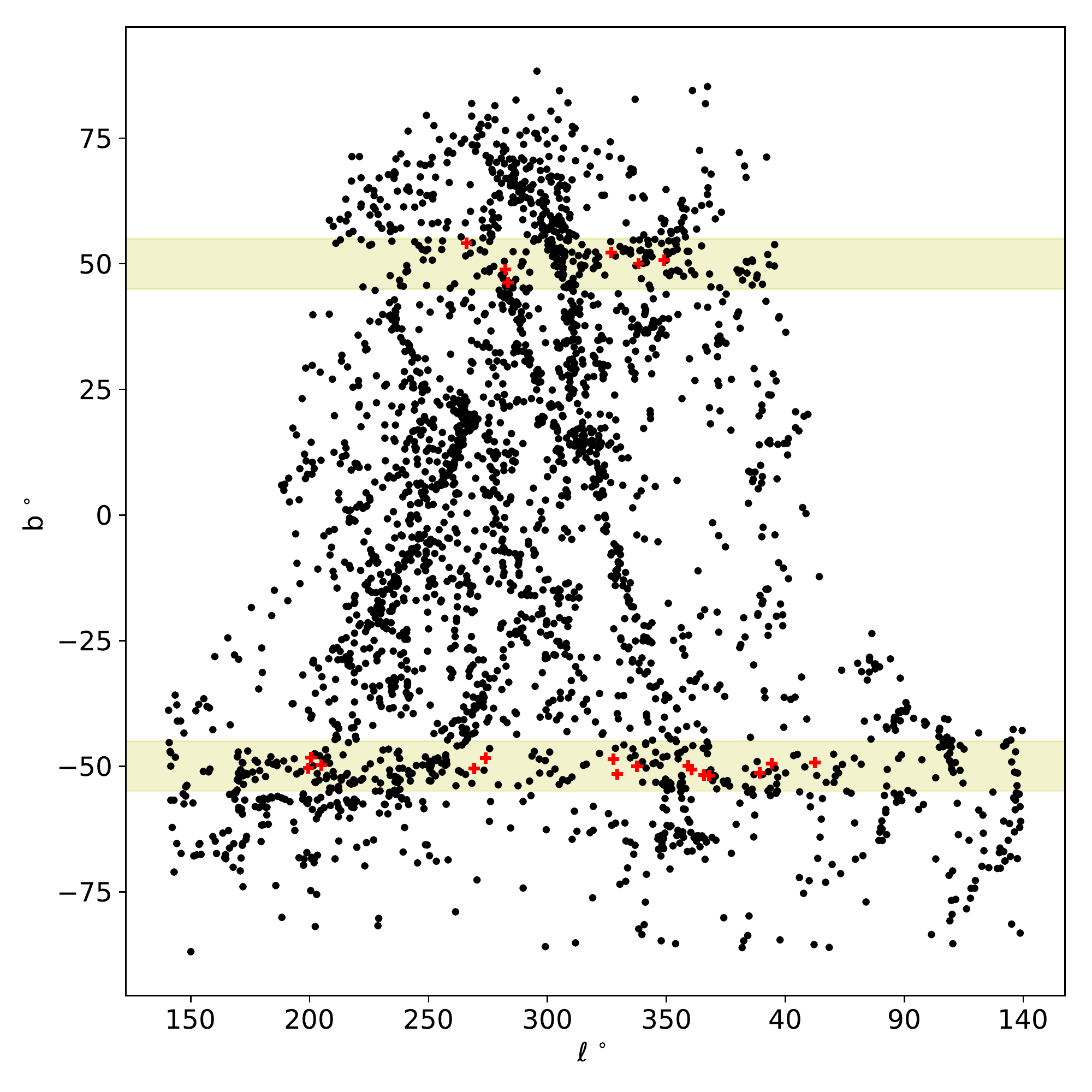}
\end{minipage}
  \caption{Comparison of galaxy locations (black dots) with ASKAP FRB locations (red crosses) for the 2MRS (top panel) and HIPASS (bottom panel) surveys.  Galaxies are shown with redshifts between $z\sim0.002-0.01$, which is the FWHM of the parameterized redshift bin centered at $\langle z \rangle \sim 0.007$.  The yellow bands show the ASKAP FRB survey region of $45^\circ<|b|<55^\circ$.}
  \label{fig:map}
  \end{figure} 
  
  \begin{figure}
    \begin{minipage}{\linewidth}
        \center
    \includegraphics[width=\textwidth]{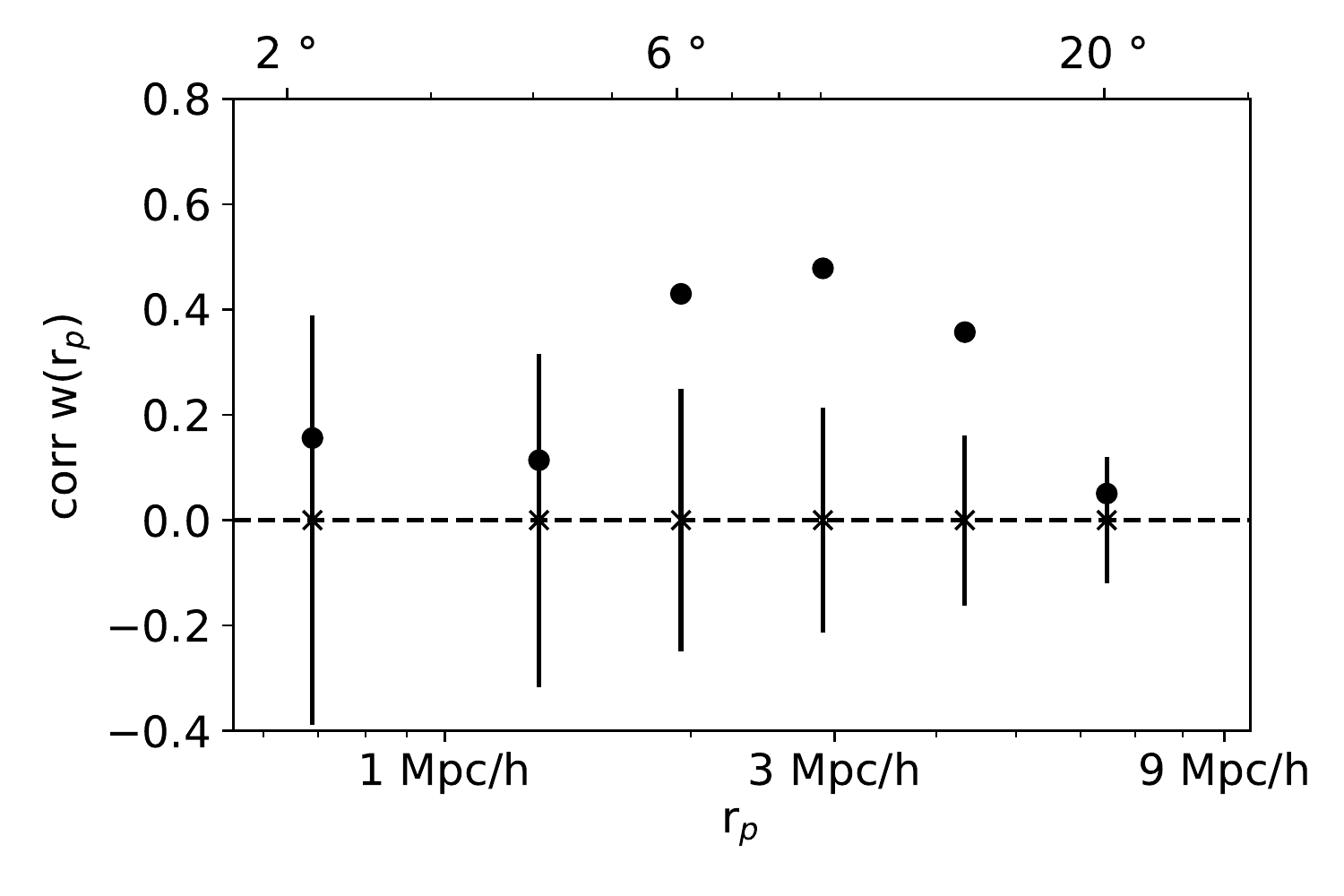}
    \vspace{-0.7cm}
    \vfill
    \includegraphics[width=\textwidth]{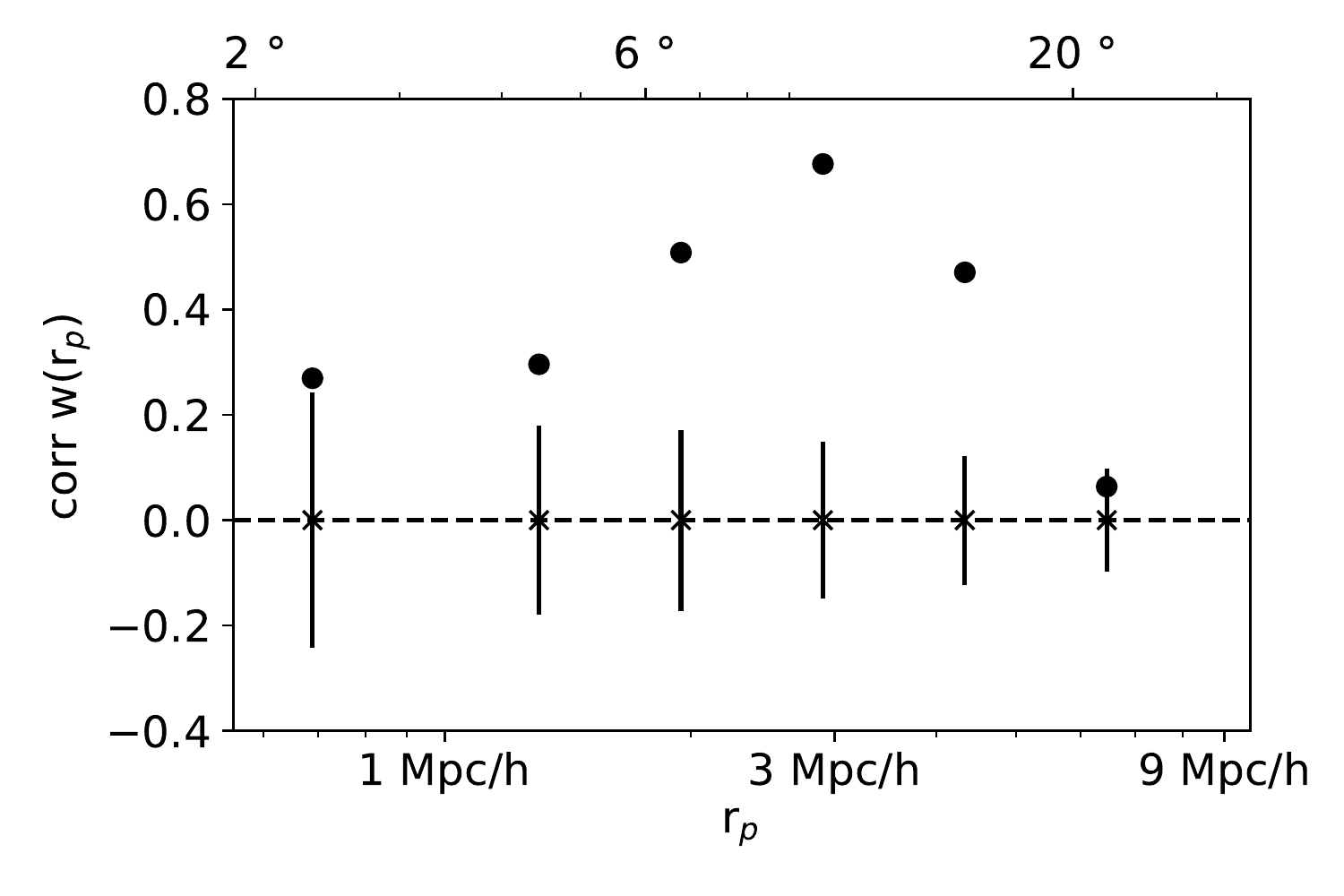}
    \end{minipage}
     \caption{(Dots) the measured correlation between the ASKAP FRBs and 2MRS (top) and HIPASS (bottom) as a function of projected distance, using galaxies weighted by Eq. \ref{eq:denZ} with $\langle z \rangle=0.007$.  
     The error bars show 1$\,\sigma$ bootstrap errors generated using randomized FRB locations. 
    } 
    \label{fig:corr}
   \end{figure}
\section{FRB galaxy cross-correlation}
\label{sec:corr}
We now have two arguments that FRBs may come from lower redshifts than their DMs might suggest, or equivalently that the host contribution to the DM is non-negligible.  This motivates an attempt to see if the locations of FRBs are correlated with large-scale structure in the nearby universe.  We will make use of the clustering redshift technique \citep{13Menard}, which looks for large-scale spatial correlation with a tracer population of known redshift.

%Clustering-based redshift estimation
\subsection{correlation estimator}
To use the clustering redshift technique, we correlate the positions on the sky of our FRB sample with a three-dimensional map of galaxy positions. We apply a redshift weight $P(z,z_0)$ to the galaxy distribution:
\begin{align}
    N_G(\mathbf{\theta},z_0)=\int_z n_G(\mathbf{\theta},z)P(z,z_0)\,dz,
\end{align}
where $n_G(\mathbf{\theta},z)$ is the number density of galaxies at redshift z. By weighting in this manner, we have isolated galaxies in some redshift range set by $z_0$. 

The simplest form of $P(z,z_0)$ will be a top-hat function centered at $z_0$, as used in, e.g.\,\citet{15Rahman}. In this case, the approach is equivalent to dividing the galaxies into discrete redshift bins. If the cross-correlation S/N in each bin is high enough, the FRB redshift distribution can be approximated as proportional to the correlation amplitudes. In our case however, the available number of FRBs is small enough that any individual thin top-hat bin will be severely noise-dominated.  We could widen the bins in an attempt to increase signal-to-noise, but combining the freedom to choose the exact bin width with a low-significance expected signal makes this analysis vulnerable to the look-elsewhere effect (even if there was no large-scale correlation, we could find a high-significance result by sheer coincidence due to the large parameter space explored). In addition, as the width of the top-hat function approaches the width of the source distribution, it is no longer reasonable to assume that the sources are uniformly distributed in the redshift bin. In this case, the top-hat weighting scheme will be sub-optimal.

We therefore adopt a smoothly varying redshift weight without hard boundaries, 
 \begin{equation}
P(z,z_0)=\frac{\beta}{z_0\Gamma(1+\frac{2}{\beta})}\,
\Big(\frac{z}{z_0}\Big)^2\, \exp \bigg[ -\Big(\frac{z}{z_0}\Big)^\beta \bigg]\,,
\label{eq:denZ}	
\end{equation}
This function is widely used in gravitational lensing studies to assemble the redshift density distribution of background sources (eg.\,\citealt{02Weinberg,03Pen}). The $z/z_0^2$ term comes from the fact that in the limit of very small redshift, the universe is Euclidean; the exponential term describes the decrease in number counts at high-$z$ as faint sources fall below the detection threshold. We use $\beta=1.5$, which is the value that fits well for many optical/infrared galaxy surveys (eg. \citealt{CFRS95Lilly,2MRS}). This can be sub-optimal if the FRB luminosity function behaves dramatically differently from that of galaxies. However, it satisfies the conditions we discussed above, as it selects wide redshift bins with a weighting scheme that likely at least roughly matches the true redshift distribution, and the choice of the width is not fine-tuned\footnote{We have also perfomed the analysis with top-hat bins of various widths, and we see qualitative evidence for a peak in the clustering amplitude near what we obtain with Eq. \ref{eq:denZ}. We however feel the above weighting scheme provides a more robust estimate of the significance of our result.}. With the chosen value of $\beta$, the mean redshift for a distribution weighted by Eq. \ref{eq:denZ} will be $\langle z \rangle \approx 1.5 z_0$. 

We estimate the correlation between the FRB locations and the weighted galaxy field with an estimator equivalent to the one defined in \citealt{83Davis}:
\begin{equation}
w(r, \azm)=\frac{D_\mr{F}D_\mr{G}(r, \azm)}{\langle R_\mr{F}D_\mr{G}(r, \azm)\rangle}-1	
%=\sum_i^N  D_G(F_i,r)/\langle \sum_i^N D_G(R_i,r) \rangle -1
\end{equation}
$D_\mr{F}D_\mr{G}(r,z_0)$ is the average galaxy number density at a given distance $r$ from FRBs: 
\begin{equation}
    D_\mr{F}D_\mr{G}(r,z_0)=
    \frac{1}{N}\sum_{i=0}^N\,N_G^i(r,z_0),
\end{equation}
where $N_G^i(r,z_0)$ is the number density of the weighted galaxy population at projected separation r from the ith FRB. The $D_\mr{F}D_\mr{G}(r,z_0)$ will be above average if FRBs are correlated with the large scale structures at the selected redshift range. 
 $R_\mr{F}D_\mr{G}(r)$ is the same quantity computed for randomized FRB locations. The expectation value of the denominator is the same as the mean galaxy number density. 

We quote uncertainties on the power spectrum using the bootstrap error computed from our randomly generated FRB catalogs,
\begin{equation}
	\sigma_{w(r, \azm)}=\sigma_{R_\mr{F}D_\mr{G}(r, \azm)}/\langle R_\mr{F}D_\mr{G}(r, \azm)\rangle,
\end{equation}
where $\sigma_{R_\mr{F}D_\mr{G}(r, \azm)}$ is the standard deviation of the number density of galaxies near randomized FRB locations. 

By varying $z_0$, we see how the correlation varies with galaxy redshift. We compare the significance of cross-correlation of different redshifts with the integrated cross-correlation: 
\begin{equation}
	\overline{w}(\azm)=\int_{r_\mr{min}}^{r_\mr{max}}
	dr\, W(r, \azm) w(r, \azm), 
\end{equation}
with a weight function:
\begin{equation}
	W(r, \azm)\propto\frac{r^{-\gamma}}{\sigma_{w(r, \azm)}^2}.
	%{\int_{r_\mr{min}}^{r_\mr{max}} \frac{r^{-\gamma}}{\sigma_{w(r, \azm)}^2}\,dr},
	\label{eq:weight}
\end{equation}
The $r^{-\gamma}$ factor weights the correlation function with the power law spatial scaling generally expected for cosmological density fields \citep{13Menard}. We choose the value of $\gamma$ to match that computed from the auto-correlation functions of our galaxy samples. If the FRB distribution traces the same structure as the galaxy distributions, the cross-correlation will have the same $r$ dependence as the auto-correlation. 
Correlations at small r are heavily weighted by the $r^{-\gamma}$ , however, they also have large statistical fluctuations due to the smaller sample size. 
We additionally weight each r by its inverse variance $1/\sigma^2$.
%Due to the relatively low S/N at small r due to smaller sample size, we additionally weight each r with $1/\sigma^2$.  
%The S/N of the weighted cross-correlation shows the significance of fitting the cross-correlation with $r^[-\gamma}$. 
% of 2MRS survey .

\subsection{ASKAP-2MRS correlation}
The 2MASS Redshift Survey (2MRS) \citep{2MRS} includes a sample of 44,599 galaxies with $K_s \leq 11.75$ mag and galactic latitude |b| $\geq 5^\circ$ ($\geq 8^\circ$ toward the Galactic bulge), which is 97.6\% complete and covers 91\% of the sky. 
The detected galaxy redshift distribution can be well-fitted by Eq. \ref{eq:denZ} with $z_0\sim0.02$ and \az$\sim0.03$. 
The 2MRS galaxy auto-correlation function is proportional to $r^{-0.9}$ \citep{18Krumpe}, so we set $\gamma=-0.9$ in Eq. \ref{eq:weight}. 
%There are $\sim 5000$ galaxies with $z>0.05$. %, which are dense enough to indicate the large scale structure. 
We vary $z_0$ between 0.003 and 0.03 in Eq. \ref{eq:denZ} to study how the correlation amplitude $\overline{w}(\azm)$ scales with redshift. 
The result is shown in Figure \ref{fig:totcorr}.  The statistical significance peaks at 3\,$\sigma$ at \az$ \sim0.007$. The full estimated $w(r,z_0=0.007)$ is shown in the upper panel of Figure \ref{fig:corr}.
This is consistent with the expected \az$ \sim0.01$ we estimated in Table \ref{tab:dm} by comparing these ASKAP FRBs to those in CHIME.  It is a somewhat lower redshift than we obtained by comparing the ASKAP and Parkes FRBs, but as stated above we consider the Parkes/ASKAP number to be an upper limit.

For illustration purpose, figure\,\ref{fig:map} overplots the location of ASKAP FRBs and the 2MRS galaxies for $0.002<z<0.01$. 
The redshift boundaries are selected to be the half maximum of the weighting function with \az$ = 0.007$.

\subsection{ASKAP-HIPASS correlation}
Given the low mean redshift we obtain with the ASKAP/2MRS correlation, we can test the robustness of our result by carrying out a similar analysis with the H\,I Parkes All Sky Survey (HIPASS).  The HIPASS catalogue contains 5317 galaxies observed in the 21cm spin-flip line between redshifts $\sim0$ and 0.04 \citep{04Meyer,06Wong}.  Eq. \ref{eq:denZ} with $z_0\sim0.006$ approximates the redshift distribution of HIPASS sources, though the fit is not as good as for the ASKAP sources.

HIPASS covers 71\% of the sky with declination range $-90^\circ<\delta<+25^\circ$. 
Studies on the Southern sky catalogue reveal it to be 99\,\% complete at a peak flux of 84 mJy and an integrated flux of 9.4 Jy\,km/s \citep{04Zwaan}.  
As HIPASS galaxies are selected on their 21 cm emission, they will have different correlation statistics than the 2MRS galaxies, which are selected in the near infrared.  The projected HIPASS auto-correlation is proportional to $r^{-0.6}$ \citep{11Passmoor} as opposed to $r^{-0.9}$ for 2MRS galaxies\,\citep{18Krumpe}. 

We again vary $z_0$ in Eq. \ref{eq:denZ} and 
plot the measured weighted cross-correlations between the weighted FRBs and the HIPASS galaxies in Figure\,\ref{fig:totcorr}. We only use $z_0$ between 0.003 to 0.009, as HIPASS spans a smaller redshift range than 2MRS.
The highest correlation again appears at $\langle z\rangle\sim0.007$ with $5\sigma$ significance.  
Unfortunately, we are unable to verify the lack of correlation at $z\gtrsim0.02$ seen in ASKAP$\times$2MRS, due to the limited sample from HIPASS at higher redshift.
%we are unable to see whether the correlation exists at $\langle z\rangle \sim 0.07$ or 0.3.
The correlation function at this redshift is shown in Figure\,\ref{fig:corr} lower panel. The measured correlation function is noisy, but within $1\sigma$ of the predicted $r^{0.6}$ scaling.

%add a summary
In summary, we cross-correlate the location of ASKAP FRBs with two galaxy surveys of different tracers, 2MRS and HIPASS, and see that both correlations peak at $z\sim 0.007$ with 3\,$\sigma$ and 5\,$\sigma$ significance respectively. 
The current large sky redshift survey depth limits our ability to look for correlations at higher redshift, such as $z\sim0.3$ predicted by the IGM dominated DM assumption. With future large-scale structure experiments such as 21\,cm intensity mapping surveys \citep{CHIME,Tianlai,HIRAX}, the evidence will be more complete. 

\section{Conclusion}
\label{sec:conclude}
In this paper, we perform three statistical tests to infer the distances of ASKAP FRBs. 
We show that the number of events of similar luminosity in ASKAP does not scale properly with DM cube when attributing all the excess DMs (net DM minus the contribution from the Milky Way) to the IGM. 
It suggests that a noticeable fraction of the excess DM should come from sources other than the IGM.
%These results indicate that most of the DM is intrinsic to the host. 
Furthermore, by comparing the average DMs of FRBs from CMIME and ASKAP, we infer an average redshift of $Z_\mr{ASKAP}\sim 0.01$.  A similar comparison between Parkes and ASKAP yields $z_\mr{ASKAP}\sim0.07$, which is expected to be biased high due to the different RFI environment. 
Both values are much smaller than the upper limit $z_\mr{ASKAP}\sim0.3$ inferred from attributing the excess DM entirely to the IGM.

To further constrain the distances, we cross-correlate the locations of the ASKAP FRBs with existing large-sky redshift surveys. 
We obtain a 3\,$\sigma$ correlation with 2MRS and a 5\,$\sigma$ correlation with HIPASS at $z\sim0.007$. 
These results are consistent with the redshift of the most likely host galaxy of ASKAP FRB 171020 \citep{Mahony2018AFRB171020}, which is found at 
$z=0.00867$. 

Results from all three tests along with the location of FRB 171020 suggest that the origins of the ASKAP FRBs are significantly closer than the distance estimated from the excess DM, and a considerable amount of the DM is intrinsic to the host galaxies. 
Although each individual test does not stand on its own, the fact that several independent tests point to the same conclusion lends credibility to the result.
%The result suggest a DM of several hundred from the FRB host galaxies, which is difficult to explain with interstellar medium (ISM) contribution through a random line of sight in the host galaxy. 
%A special local environment of accumulated electron density will be preferred. 

The inferred proximity of the FRBs suggests that the energy estimated from the direct DM/distance conversion as in \citep{Shannon2018TheSurvey} may be off by orders of magnitude. 
It also indicates that the number of bright events may be considerably less than previously expected, which suggests an underlying luminosity function with a steep tail at the bright end. This will influence our view of detection rate for high redshift events.  
Meanwhile, our results suggest that more care should be taken when considering  FRBs as a probe of cosmological parameters, as one may not be able to simply use DM as a proxy for distance. 
%All those aspects could be further addressed as the number of detected FRBs increases and more localization is performed. 

These results also suggest new 
%The new insight from this work provides us with new 
constraints on theoretical models. We found that a large portion of the DM,  typically several hundred $\rm pc \cdot cm^{-3}$, is intrinsic to the host. It is larger by an order of magnitude than most lines of sight through our galaxy \citep{Cordes2003NE2001.Electrons}. Therefore, any complete theoretical description for FRBs must also explain the DM. So far, most theoretical models attempt to explain different aspects of FRBs individually, such as the rate, coherent emission, luminosity and duration. One model that does account for the DM proposes FRBs come from young magnetars, and the dispersion measure comes from the surrounding supernova remnant \citep{Margalit2018ANebula, Metzger2019FastWaves}. Another possibility is that FRBs happen in environments where the ambient medium is considerably denser than the ISM, such as galactic centres \citep[e.g.][]{Thompson2017GiantDipoles, Thompson2017TinyExplosions}. These are just two examples, and we defer to future works the task of determining the favoured models. 
With the upcoming large number of detected FRBs, the power of statistical arguments like these will dramatically increase, enabling precision studies of the nature of FRBs.

\section*{Acknowledgements}
We acknowledge Ue-Li Pen for suggesting this interesting topic, Marten van Kerkwijk for encouraging the write up of the work, and Mubdi Rahman for useful discussions.
%%%%%%%%%%%%%%%%%%%%%%%%%%%%%%%%%%%%%%%%%%%%%%%%%%



\bibliographystyle{mnras}
\bibliography{askap,almog} % if your bibtex file is called example.bib
%\bibliographystyle{mnras}
%\bibliography{almog}

%%%%%%%%%%%%%%%%%%%%%%%%%%%%%%%%%%%%%%%%%%%%%%%%%%

%%%%%%%%%%%%%%%%% APPENDICES %%%%%%%%%%%%%%%%%%%%%
\appendix
\section{}
\label{app:lum}
Assume a source has a luminosity function of $f(L)$. 
The minimum flux a detector A can detect is $S_m$. 
If the detection of the source is flux rather than volume limited, in the nearby universe where the Euclidean assumption holds, the faintest source the detector can probe at distance r should have a luminosity of $L_m(r)=4\pi r^2 S_m$. 
Therefore, the number of sources visible to A at a distance r will be: 
\begin{equation}
	N(r)=4\pi r^2 \int_{4\pi r^2 S_m}^{+\infty}  f(L)\,dL
\end{equation}
The average distance of all the sources detected by A will be: 
\begin{align}
		\langle r\rangle &=\frac{\int r N(r)\,dr}{\int N(r)\,dr} \\
		&=\frac{\int 4\pi r^3 \int_{4\pi r^2 S_m}^{+\infty}  f(L)\,dL dr}
		{\int 4\pi r^2 \int_{4\pi r^2 S_m}^{+\infty}  f(L)\,dL dr}
\end{align}

Assume detector B is K times more sensitive than detector A, and therefore can detect a minimum flux of $S_m^\prime=S_m/K$. 
Then the average distance of the sources detected by B will be: 
\begin{align}
		\langle r^\prime\rangle
		=\frac{\int 4\pi r^3 \int_{4\pi r^2 S_m/K}^{+\infty}  f(L)\,dL\,dr}
		{\int 4\pi r^2 \int_{4\pi r^2 S_m/K}^{+\infty}  f(L)\,dL\,dr}
\end{align} 
Assume $r^\prime=r/\sqrt{K}$,
\begin{align}
		\langle r^\prime\rangle
		%&=\frac{\int 4\pi {(\sqrt{K}r^{\prime})^3 \int_{4\pi r^{\prime2} S_m}^{+\infty}  f(L)\,dL\,d\sqrt{K}r^\prime}}
		%{\int 4\pi {(\sqrt{K}r^{\prime})^2 \int_{4\pi r^{\prime2} S_m}^{+\infty}  f(L)\,dL\,d\sqrt{K}r^\prime}} \\
		&=\sqrt{K}\frac{\int 4\pi r^{\prime3} \int_{4\pi r^{\prime2} S_m/K}^{+\infty}  f(L)\,dL\,dr^\prime}
		{\int 4\pi r^{\prime2} \int_{4\pi r^{\prime2} S_m/K}^{+\infty}  f(L)\,dL\,dr^\prime} \\
		&=\sqrt{K}\,\langle r\rangle
\end{align} 

Therefore, the ratio of the average distance of the sources detected by detector A and B is proportional to the square root of the relative sensitivity of the two instrument, despite the detailed form of the luminosity function: 
\begin{equation}
	\frac{\langle r^\prime\rangle }{\langle r\rangle }=
	\sqrt{\frac{S_m}{S_m^\prime}}
\end{equation}

When the redshift is low, the Hubble constant could be consider as a constant, we have 
\begin{equation}
	\frac{\langle z^\prime \rangle}{\langle z\rangle }=
	\sqrt{\frac{S_m}{S_m^\prime}}
\end{equation}
where $\langle z\rangle$ is the average redshift of the detected sources.

% Don't change these lines
\bsp	% typesetting comment
\label{lastpage}
\end{CJK*}
\end{document}